\documentclass[reqno, 11pt]{amsart}

\numberwithin{equation}{section}

\usepackage{bbm}
\usepackage{amssymb, amsmath, hyperref}
\usepackage{enumerate}

\newcommand{\tfix}{T_{\mathrm o}}
\newcommand{\tlb}{T_{\text{\sc l}}}
\newcommand{\trb}{T_{\text{\sc r}}}
\newcommand{\betal}{\beta_{\text{\sc l}}}
\newcommand{\betar}{\beta_{\text{\sc r}}}
\newcommand{\partiallb}{{\partial_{\text{\sc l}}}}
\newcommand{\partialrb}{{\partial_{\text{\sc r}}}}
\newcommand{\tscprof}{\mathbf{T}^{\mathrm{sc}}}
\newcommand{\gsc}{{\mathrm g}^{\mathrm{sc}}}
\newcommand{\bfgsc}{{\mathbf g}^{\mathrm{sc}}}
\newcommand{\tsc}{T^{\mathrm{sc}}}
\newcommand{\tlin}{T^{\beta\mathrm{lin}}}
\newcommand{\vmin}{v^{({\rm min})}}
\newcommand{\ppr}[1]{{[#1]}}
\newcommand{\set}[1]{\{ #1 \}}
\newcommand{\norm}[1]{\Vert #1 \Vert}
\newcommand{\mean}[1]{\langle #1 \rangle}
\newcommand{\dmean}[1]{\langle\langle #1 \rangle \rangle}
\newcommand{\rmd}{{\rm d}}
\newcommand{\rme}{{\rm e}}
\newcommand{\Z}{{\mathbb Z}}
\newcommand{\R}{{\mathbb R}}

\newcommand{\vep}{\varepsilon}

\newcommand{\bv}{e}
\newcommand{\en}{\mathcal{E}}

 \newtheorem{theo}{Theorem}
 \newtheorem{lemma}{Lemma}

\begin{document}

\title[Anharmonic crystals with self-consistent reservoirs]{Heat Conduction
  and Entropy Production in Anharmonic Crystals with Self-Consistent Stochastic
  Reservoirs}

\author{F.~Bonetto}
\address{Federico Bonetto\\ School of Mathematics\\ 
Georgia Institute of Technology\\ Atlanta, GA 30332\\USA}
\email{federico.bonetto@math.gatech.edu}
\author{J.~L.~Lebowitz}
\address{Joel L. Lebowitz\\ Department of Mathematics and Physics\\ Rutgers University\\
Piscataway, NJ\\ USA}
\email{lebowitz@math.rutgers.edu}
  \author{J.~Lukkarinen}
\address{Jani Lukkarinen\\ Zentrum Mathematik\\ 
Technische Universit\"{a}t M\"{u}nchen\\ 
Boltzmannstr. 3\\
85747 Garching\\ 
Germany\\
{\em and\/}\\
Department of Mathematics and Statistics\\
University of Helsinki\\
P.O.~Box 68\\
00014~Helsingin yliopisto\\
Finland
}
\email{jani.lukkarinen@helsinki.fi}
\author{S.~Olla}
\address{Stefano Olla\\ Ceremade, UMR CNRS 7534\\
Universit\'e de Paris Dauphine\\
Place du Mar\'echal De Lattre De Tassigny\\
75775 Paris Cedex 16\\France
}
\email{olla@ceremade.dauphine.fr}
\urladdr{http://www.ceremade.dauphine.fr/\~{}olla}

\begin{abstract}
  We investigate a class of anharmonic crystals in $d$ dimensions,
  $d\ge 1$, coupled to both external and internal heat baths of the
  Ornstein-Uhlenbeck type. The external heat baths, applied at the
  boundaries in the $1$-direction, are at specified,
  unequal, temperatures  $\tlb$ and $\trb$.  The temperatures of the internal
  baths are determined in a self-consistent way by the requirement
  that there be no net energy exchange with the system in the
  non-equilibrium stationary state (NESS). We prove the existence of
  such a stationary self-consistent profile of temperatures for a finite
  system and show that it minimizes the entropy production to leading order
  in $(\tlb -\trb)$. 
  In the NESS the heat conductivity $\kappa$ is defined as the heat flux per
  unit area divided by the 
  length of the system and $(\tlb -\trb)$.
  In the limit when the temperatures of the
  external reservoirs go to the same temperature $T$, 
  $\kappa(T)$ is given by the Green-Kubo
  formula, evaluated in an equilibrium system coupled to reservoirs all
  having the temperature $T$. 
  This $\kappa(T)$ remains bounded as the
  size of the system goes to infinity. 
  We also show that the corresponding
  infinite system Green-Kubo formula yields a finite result. 
  Stronger results are obtained under the assumption that the 
  self-consistent profile remains bounded.
\end{abstract}
 
\date{\today}

\thanks{ }

\maketitle

\vspace*{-1.0em}
\begin{center}
{\em Dedicated to J\"urg Fr\"ohlich and Tom Spencer with friendship and appreciation}
\end{center}
\vspace*{0em}

\section{Introduction}
\label{sec:intro}

The rigorous derivation of Fourier's law of heat conduction for
classical systems with Hamiltonian bulk dynamics 
(or for quantum systems with Schr\"{o}dinger evolution) with boundaries kept
at different temperatures is an
open problem in mathematical physics \cite{blr}.  The
situation is different for systems with purely stochastic dynamics,
e.g.\ for the Kipnis, Marchioro, Presutti (KMP) model \cite{KMP82},
where such results can be readily derived \cite{KL99,spohn91}.  An interesting
area of current research are hybrid models in which the time evolution
is governed by a combination of deterministic and stochastic dynamics.
The deterministic part of the dynamics is given by the usual 
Hamiltonian evolution.
The stochastic part can be of two different types.  In the first
type, the stochastic part is constructed to strictly 
conserve the energy, as studied in \cite{bo}, or conserve also momentum, 
as in \cite{bborev, bbo}.
In the second type, studied in \cite{brv} and \cite{bll}, the
stochastic part is implemented by
coupling the particles of the system to ``internal'' heat baths with
which they can exchange energy. To obtain a heat flow between external
reservoirs at specified temperatures $\tlb, \trb$, acting at the left and
right boundaries of the system,
the temperatures of the internal
heat baths are chosen in a self-consistent manner by the requirement
that in the nonequilibrium stationary state (NESS) there be no net
energy flux between these baths and the system \cite{brv, bll}.
Because of this self-consistency condition, there is an average constant
energy flux across the system in the 
NESS, supplied by the external reservoirs at specified, unequal,
temperatures coupled to the boundaries of the system, and then carried
by the Hamiltonian dynamics. A proof of Fourier's law for both types
of hybrid models has been obtained for the case when the Hamiltonian
dynamics is linear \cite{bo,bll}, i.e., for a system of coupled
harmonic oscillators.

In the present work we investigate the self-consistent model for anharmonic
crystals.  Unlike the case of the harmonic system, where it is known
that Fourier's law does not hold when the ``noise'' is turned off (the
heat conductivity then becoming infinite), one expects that in the
anharmonic system with a pinning self-potential the
conductivity will stay finite, i.e., it will satisfy Fourier's law,
even when the strength of the noise goes to zero.  We are quite far from
proving this, however.  What we do show here is that,
for these anharmonic systems, conductivity for the finite system, defined
by first letting both $\tlb$ and $\trb$ approach the same value, 
is given by a Green-Kubo
formula.   We also prove that this Green-Kubo
conductivity is bounded in the system size, whenever the noise is finite.

These results are obtained by studying the entropy production in the 
reservoirs in the NESS specified by the temperatures of all the reservoirs.
We prove that the self-consistent profile minimizes, among
all possible temperature profiles, the entropy production to 
the leading order in the difference of the boundary temperatures
$\tlb-\trb$.  
We then prove a uniform bound for the entropy production
of a stationary state with a profile linear in the inverse
temperatures. 
This leads to a bound on the leading
term of the conductivity of the self-consistent system, given by the
Green-Kubo formula for the finite system with all reservoirs at the
same temperature $T$.  

Furthermore, we show that the corresponding Green-Kubo formula for the
infinite system, giving the conductivity of
the infinite system as a space-time integral of the energy-current
correlations, is convergent.  The bound we derive implies that the conductivity vanishes in the limit of infinitely strong coupling to the reservoirs.  This behavior is also apparent in the explicit expression of the conductivity of the corresponding harmonic system (see equation (7.10) in \cite{bll}).
The violent contact with the reservoirs most likely makes local equilibrium so
strong that eventually no transmission is possible.

There are no comparable results for anharmonic crystals with the 
first type of hybrid dynamics, but only some bounds on the conductivity 
\cite{bbo}.
Under the assumption that the self-consistent temperature profile remains
bounded, we show that 
the conductivity of the finite systems  with a fixed 
$\tlb-\trb>0$ is uniformly bounded in the size of
the system. (This assumption is ``clearly'' correct but we are unable
to prove it, see section 9.)

The model considered is described in section \ref{sec:model} while
section \ref{sec:summary-results} contains a summary of the results
proven in this paper. The
existence of a NESS with a self-consistent temperature profile is
proven in section \ref{sec:self-cons-prof}. Entropy production in the
NESS is discussed in section \ref{sec:entropyprod}, and in section
\ref{sec:temp} we prove that the stationary state corresponding to the
self-consistent profile minimizes, at the leading order in the
temperature difference $(\tlb-\trb)$, the entropy production.  Thermal conductivity in
the NESS is discussed in section \ref{sec:cond} and for the infinite
homogeneous system in section \ref{sec:cond-infin-syst}.
Finally, in Section \ref{sec:conclusions} we present some concluding remarks.

\section{Time Evolution}
\label{sec:model}

Atoms are labeled by $x=(x_1,\ldots,x_d) \in \{-N,\dots, N\}^d = 
\Lambda_N$, $N\ge 1$. Each atom is in
contact with a heat reservoir at temperature $T_x$.
The interactions with the reservoirs
are modeled by Ornstein-Uhlenbeck processes at corresponding
temperatures. The atoms have all the same mass $m=1$. Their velocities 
are denoted by $p_{x}$
 and the ``positions'' by  $q_{x}$, with $q_x,p_x\in \R$.
We consider a mixture of fixed and periodic boundary conditions.
The fixed boundary conditions are applied in the $1$-direction, and 
the corresponding boundary sites will be used to make contact with
external heat reservoirs.  In the remaining directions, we apply periodic
boundary conditions.  Explicitly, let $\overline{\partial}\Lambda_N$
denote the set with $|x_1|=N+1$ and let
$\ppr{x}_i=-N+(x_i+N) \bmod (2 N+1)$, for $i\ge 2$.
The boundary conditions are then
$q_{x}=0$, for $x\in \overline{\partial} \Lambda_N$.
In addition, we let the inner boundary of $\Lambda_N$ consist of those
$x$ with $|x_1|=N$, and we denote it by $\partial\Lambda_N$. 

As we will show, the heat flux in the stationary state will be entirely in the
$1$-direction and the properties of the system will be uniform in the $d-1$
periodic directions.
We define 
$\Lambda_N' = \{x \in\Lambda_N : -N\le x_1 < N\}$ to label
the bonds in the $1$-direction.

The Hamiltonian of the system is given by
\begin{equation}
  \begin{split}
     \label{eq:1}
  \mathcal H_N & = \sum_{x\in \Lambda_N} \en_x, \\
\en_x & = \frac{ p_x^2}{2}  + \sum_{j=1}^d \frac{V(q_x- q_{x-\bv_j}) +
    V(q_{x+\bv_j}- q_x)}2 + W(q_x), \qquad x \in\Lambda_N \, ,
  \end{split}
\end{equation}
where the $\bv_i$, $i=1,\ldots,d$, denote the Cartesian basis vectors.
We assume that $V$ and $W$ are smooth positive symmetric 
functions on $\mathbb R$
with quadratic growth at infinity:
\begin{equation}
\lim_{\lambda \to \infty}W''(\pm\lambda) =
  W''_\infty > 0,\quad \lim_{\lambda\to\infty}V''(\pm\lambda) = V''_\infty
  > 0 \, .
\label{eq:32}   
\end{equation}
Clearly then, there are $C_1,C_2>0$ such that
\begin{align} 
 &  C_1 (q^2-1) \le V(q) \le C_2 (q^2+1), \ 
  C_1 (q^2-1) \le W(q) \le C_2(q^2+1)\, .
\end{align}

The dynamics is described by the following system of stochastic
differential equations:
\begin{equation}
  \label{eq:2}
  \begin{split}
    dq_x &= p_x dt\, , \\ 
    dp_x &= -{\partial_{q_x} \mathcal H_N}\;  dt -\gamma_x p_x \; dt +
    \sqrt{2 \gamma_x T_x}  \; dw_x(t) \, ,
  \end{split}
\end{equation}
with $\gamma_x>0$ for all $x \in\Lambda_N$. 
Here $w_{x}(t), x \in \Lambda_N$, are independent standard Brownian
motions (with $0$ average and diffusion equal to 1).
The generator of this process has the form
\begin{equation}
  \label{eq:3}
  \begin{split}
    L_N & = \sum_{x \in \Lambda_N} \left({\partial_{p_x} \mathcal H_N}
    \partial_{q_x} - {\partial_{q_x} \mathcal H_N}
    \partial_{p_x}\right) +   \sum_{x\in \Lambda_N} \gamma_x \left(T_x
    \partial_{p_x}^2 - p_x \partial_{p_x} \right) \\
   & =  A + S\, ,
  \end{split}
\end{equation}
where $A$ is the Hamiltonian part, anti-symmetric in the momentum variables,
and $S$ is the symmetric part corresponding to the action of the reservoirs.
Then
\begin{equation}
  \label{eq:5}
  L_N \en_x = 
  \sum_{i=1}^d \left( j_{x-\bv_i,x} -  j_{x, x+\bv_i} \right) +  J_x, \qquad
  x\in \Lambda_N 
\end{equation}
with $J_x=\gamma_x (T_x - p_x^2)$ and 
\begin{align}
  \label{eq:6}
   &  j_{x,x + \bv_i} = 0, \qquad \text{if } \ppr{x}\not \in \Lambda_N  
   \text{ or } \ppr{x+\bv_i}  \not \in \Lambda_N ,  \\
  &   j_{x,x + \bv_i} = -\frac 12 (p_{\ppr{x}} + p_{\ppr{x+\bv_i}})
  V'(q_{\ppr{x+\bv_i}} - q_{\ppr{x}}), \qquad \text{otherwise} \, .   \label{eq:6b}
\end{align}
In particular, then $j_{x,x + \bv_1}$ can be non-zero only if $x\in \Lambda'_N$.

In section 3 of \cite{msh} it is shown that, for any choice of the
temperatures $\mathbf T = \{T_x\ge 0\}$, 
  there exists an explicit
 Lyapunov function for the corresponding stochastic evolution, as long
 as $\gamma_x >0$ for all $x$. This implies the existence of the
 corresponding stationary measure that we will denote by 
$\mu(\mathbf T)$. 

If at least one $T_x>0$, then the generator $L_N$ defined in (\ref{eq:3}) is
(weakly)-hypoelliptic,  
in the sense that the Lie algebra generated by the \emph{vector fields} 
$\{A, \partial_{p_x}, x\in \Lambda_N\}$ has full range in the tangent space
 of the phase space $(\R^{2 d})^{\Lambda_N}$.
In particular, the dynamics has
probability transitions with smooth densities with respect to the
Lebesgue measure on the phase space. If all $T_x >0$, also the corresponding 
control problem has a strong solution  
(cf.\ section 3 in  \cite{msh}, or \cite{Hairer08}) and
uniqueness of the stationary measure follows from these properties. 
These methods could be extended to the case $T_x \ge 0$, at least if 
$\mathcal H_N(p,q)$ is strictly convex \cite{matt-prv}. The
investigation of the uniqueness of the stationary measure goes beyond
the purposes of the present paper, in particular, since 
zero temperatures will be relevant only in
the general proof of
existence of a self-consistent temperature profile in Section
\ref{sec:self-cons-prof}.  So we will assume the uniqueness even in the
case of temperatures not strictly positive.  

The spatial periodicity will be exploited in the following to remove (most
likely irrelevant) technical difficulties associated with irregular
boundary behavior.  To this end, we will assume that also the heat bath
couplings respect this periodicity, i.e.,
{\em we will always assume that $\gamma_x$ depends only on $x_1$.}
Then in the case where also $T_x$ depends only on $x_1$, the
stochastic dynamics is 
fully invariant under periodic translations.  Since the stationary measure 
$\mu(\mathbf T)$ is unique, then also any of the corresponding
expectation values must be invariant.

We denote the constant temperature profile, $T_x = \tfix$ for all
$x\in \Lambda_N$, as $\mathbf \tfix$. 
Then $\mu(\mathbf \tfix) = \mu_{\tfix}$, the Gibbs
measure at temperature $\tfix$, defined by
\begin{equation}
\mu_{\tfix} = Z_{\tfix}^{-1} \exp(-\mathcal H_N(p,q)/\tfix) \rmd p \rmd q
=G_{\tfix}(p,q) \rmd p \rmd q   \label{eq:28}
\end{equation} 
We use $\mu_{\tfix}$ 
as a reference measure and denote
the related  expectation by  $\left<\cdot\right>_0$.

Computing the adjoint of $L_N$ with respect to the Lebesgue measure
we have\footnote{We wish to reserve the standard notation for adjoint for
  certain weighted $L^2$-spaces, to be introduced later.  Hence the notation
  $L_N^{*(1)}$ for the adjoint here.}
\begin{align} 
  L_N^{*(1)} = -A+ \sum_{x\in \Lambda_N} S_x^{*(1)}
\end{align}
where $S_x^{*(1)} = \gamma_x\left( T_x \partial_{p_x}^2 + 1 +
   p_x \partial_{p_x}\right)$.
We denote by $f_N= f_N(\mathbf T)$ the density of the stationary state
$\mu(\mathbf T)$ with respect to Lebesgue measure. This is the solution of 
$L_N^{*(1)} f_N = 0$. Due to hypoellipticity, $f_N$ is a smooth
function of $(p,q)$, and this implies also smoothness in $\mathbf T$.
To see this, note that $\partial_{T_y} f_N$ is the solution of the equation 
\begin{equation}
  L_N^{*(1)} \partial_{T_y} f_N = - \gamma_y \partial_{p_y}^2 f_N  \, .
\end{equation}
Since the right hand side is smooth in $(p,q)$, this
equation has a smooth solution, and smoothness in $\mathbf T$ follows
by a standard iteration of the argument.

\section{Summary of results}
\label{sec:summary-results}

Given the temperatures $\mathbf{\Theta}_R= \{\Theta_y\}_{y\in R}$ in a
set $R\subset \Lambda_N$, 
we say that a temperature profile $\mathbf T =\{T_x\}_{x\in\Lambda_N}$ is 
{\em self-consistent\/}, if $T_x=\Theta_x$ for all $x\in R$, and
the  corresponding stationary state 
has the property
\begin{equation}
  \left<p_x^2\right> = T_x, \qquad \text{for all }x\in\Lambda_N\setminus R,
\end{equation}
where $\mean{\cdot}$ denotes expectation with respect to the NESS,
$\mu(\mathbf T)$, assumed to be unique.
Eventually we may choose $R = \partial \Lambda_N$ or part of it. But
the following result is independent from the geometry. 

\begin{theo}\label{th:scexists}
  For any choice of a non-empty $R\subset \Lambda_N$,
  and for any choice of temperatures
  $\mathbf{\Theta}_R = \{\Theta_y\}_{y\in R}$ not all equal to $0$, there
  exists a self-consistent temperature profile 
  $\mathbf T =\{T_x\}_{x\in\Lambda_N \setminus R}$. 
  In addition, if $R$ and $\mathbf{\Theta}_R$
  are invariant under translations
  in all of the $d-1$ periodic directions of $\Lambda_N$, then a
  self-consistent profile invariant under these translations can be found.
\end{theo}

The main body of our results concerns the case where the reservoirs on the
two sides in the  
non-periodic direction are fixed to constant but unequal temperatures.
We call this case the {\em boundary layer setup\/}.
More explicitly, we then define
$R= \partial \Lambda_N = \partiallb \Lambda_N \cup \partialrb \Lambda_N$, 
where $\partiallb \Lambda_N = \{ x : x_1 =-N\}$ and   
$\partialrb \Lambda_N = \{ x : x_1 = N\}$, 
and we fix on the left the temperatures $T_x = \tlb$ for
$x\in \partiallb \Lambda_N$,  and on the right $T_x = \trb$ for
$x\in \partialrb \Lambda_N$, $\trb<\tlb$. 
We also set 
$\betal = \tlb^{-1}$, and $\betar = \trb^{-1}$.
Uniqueness of the self-consistent profile is not claimed in
Theorem \ref{th:scexists}, and this remains an open problem in the 
generality of the theorem. 
However, by restricting to small temperature differences and then relying
on the implicit function theorem, we can get a self-consistent profile which
is essentially unique.  
\begin{theo}\label{th:scexists2}
  For any given $\tfix>0$ and $N$, there are $\vep_0,\delta>0$ 
  with the following property: In the boundary layer setup
  with $\tlb,\trb$ such that $|\tlb-\tfix|$, $|\trb-\tfix|<\frac{1}{2}\vep_0$
  there is a self-consistent extension of the temperature profile,
  $\mathbf \tsc(\tlb,\trb)$, and the extension is unique in the sense that
  no other profile $\mathbf T$
  with $\max_x |T_x-\tfix|< \delta$ is self-consistent.
  In addition, $\mathbf \tsc$ is invariant under translations
  in all of the $d-1$ periodic directions of $\Lambda_N$, and the map 
  $(\tlb,\trb)\mapsto \mathbf \tsc(\tlb,\trb)$ is smooth.
\end{theo}

As an aside, let us remark that a careful
inspection of the proof of Theorem \ref{th:scexists2} shows that
its assumptions could be greatly relaxed, allowing for more general sets $R$
and almost arbitrary potentials $V$ and $W$.  However, since the range of
its applicability, determined by $\vep_0$, can depend on $N$ and 
might go to zero as $N\to \infty$, we have included the proof of the more
general result in Theorem \ref{th:scexists}.
Furthermore, the assumptions about the asymptotic quadratic behavior of 
$V$ and $W$ will be used in latter proofs, and thus cannot be neglected.
From now on, we assume that $\tlb-\trb$ is sufficiently small for applying
Theorem \ref{th:scexists2}, and let $\mathbf \tsc$ denote the 
corresponding self-consistent extension of the temperature profile,
which is thus  invariant under periodic translations and leads to a unique,
periodically invariant, stationary state.

For a generic profile $\mathbf T$, we define
the entropy production in a reservoir in the steady state 
$\mu(\mathbf T)$ as the energy flux entering that reservoir divided by its
temperature \cite{BL55}.  
The total steady state entropy production is then given by
\begin{equation}
  \label{eq:21}
  \sigma (\mathbf T) = \sum_{x\in\Lambda_N} \frac{\mean{-J_x}}{T_x}
  =
  \sum_{x\in\Lambda_N} \gamma_x
   \Bigl({\frac{\left<p_x^2\right>}{T_x}  - 1}\Bigr) \, .
\end{equation}
By using the local energy conservation \eqref{eq:5}
and denoting $\beta_x = T_x^{-1}$, we can write this as
\begin{equation}
  \label{eq:29a}
  \sigma (\mathbf T) = 
  \sum_{i=1}^d 
  \sum_{x\in \Lambda'_N} (\beta_{x+e_i} - \beta_{x}) \mean{j_{x,x+e_i}} \, .
\end{equation}
It is is well known \cite{BL55} that $\sigma (\mathbf T) \ge 0$. 

For the self-consistent profile $\mathbf \tsc$,
there are no fluxes to the reservoirs for $x\not\in \partial\Lambda_N$
and consequently, as will be shown below,
$\mean{j_{x,x+\bv_1}} = \bar{j}_N$ for all $x\in \Lambda'_N$. 
The entropy production \eqref{eq:29a} is then equal to 
\begin{equation}
  \label{eq:38}
   \sigma (\tscprof) = (2N+1)^{d-1} (\betar -\betal)  \bar{j}_N  \, .
\end{equation}
Thus we can estimate the magnitude of the self-consistent current by
estimating the entropy production.
\begin{theo}\label{entropybound}
\begin{equation}
  \label{eq:40}
   \sigma (\tscprof) \le (2N+1)^{d-2} (\betar -\betal)^2
   C(\mathbf \tsc, \gamma) 
\end{equation}
where, up to a constant $c$ depending only on the potentials $V$ and $W$,
\begin{equation}
  C(\mathbf \tsc, \gamma) = c \frac{\max_x \gamma_x \tsc_x}{\min_x
    \gamma_x^2} ( 1 + \max_x \tsc_x) \, . 
\end{equation}
Consequently, the average self-consistent current is bounded by
\begin{align}
0\le \bar{j}_N \le C(\mathbf \tsc, \gamma)\ 
 \frac{ \betar -\betal}{2N+1}\, .
\end{align}
\end{theo}

We expect, but are not able to prove, that 
 the self-consistent profiles remain uniformly bounded in $N$.
 From such a bound it would follow that $\bar{j}_N = \mathcal{O}(N^{-1})$.
 We expect in fact that $T_x\in [\trb,\tlb]$, as in the harmonic case
\cite{bll}, c.f., Section \ref{sec:conclusions}.  What we can
prove is that
the first order term of $\bar{j}_N$ in an expansion in the imposed temperature
gradient  
is $\mathcal{O}(N^{-1})$.  This is possible even without explicit knowledge
about the 
asymptotics of the self-consistent profile.  To this end, we 
consider also profiles $\mathbf \tlin$ which are extensions in the boundary
layer setup to a profile with linear $\beta_x$; we define
\begin{equation}
  \label{eq:44}
  (\tlin_x)^{-1} = \frac 12 \left(\frac{ \betar -\betal}{N} x_1 + \betar +
    \betal\right )  , \quad x\in \Lambda_N\, .
\end{equation}
For these profiles, the entropy production satisfies
\begin{equation}
  \label{eq:45}
  \sigma (\mathbf \tlin ) =  \frac{\betar -\betal}{2N} 
   \sum_{x\in \Lambda'_N} \mean{j_{x,x+e_i}}_{\mathbf \tlin}\, ,
\end{equation}
and we can derive a more precise bound for it.
\begin{theo}\label{lin-ep}
Given $b>0$,  there 
exists a constant $C_2(\gamma;b)$, depending only on $\gamma$, $V$, $W$,
and $b$, such that for all $\trb\le \tlb\le b$,
\begin{equation}
   \sigma (\mathbf \tlin ) \le (2N+1)^{d-2} (\betar -\betal)^2
   C_2(\gamma;b).   \label{eq:46}
 \end{equation}
\end{theo}

\medskip

Obviously, 
if $\mathbf \tfix$ is any constant temperature profile, we have
$\sigma (\mathbf \tfix) = 0$. 
Furthermore, 
$\frac{\partial \sigma}{\partial T_x}   (\mathbf \tfix) = 0$, and the second
order derivatives can also be computed, yielding the following theorem.
\begin{theo}\label{S-taylor}
  The Taylor expansion of $\sigma$ around a constant profile $\mathbf \tfix$ at
  the second order gives
    \begin{align}\label{eq:Staylor}
\sigma (\mathbf \tfix + \vep\mathbf{v}) 
= \frac{\vep^2}{\tfix^2} Q(\mathbf{v}; \tfix) + \mathcal{O}(\vep^3),
\quad  Q(\mathbf{v}; \tfix) =
\sum_{x,y\in \Lambda_N} \mathcal{J}_{y,x} v_y v_x\, ,
\end{align}
where, with
$\mean{\cdot}_0$ denoting the expectation in $\mu(\mathbf \tfix)$,
\begin{equation}
  \label{eq:1jaco}
   \mathcal{J}_{y,x} = \gamma_x
  \delta_{y,x} -   \gamma_x\gamma_{y} \mean{h_{x} (-L_N(\mathbf\tfix))^{-1} 
    h_{y}}_0,
  \quad h_x = \frac{p_x^2}{\tfix} -1\, .
\end{equation}
The matrix $\mathcal{J}$ is positive, and if $\mathcal{J}_{y,x}$ is
restricted to $x,y\in \Lambda_N\setminus \partial\Lambda_N$, it becomes 
strictly positive.
\end{theo}

We now denote $\delta T = \tlb - \trb$ and $\tfix = (\tlb + \trb)/2$.
The next result says that the self-consistent profile minimizes entropy
production, 
at least up to the leading order in the gradient of the imposed
temperature difference, $\delta T$. 

\begin{theo}\label{T-taylor}
  The self-consistent profile $\mathbf \tsc$ is a smooth function of
  $\tlb$ and $\trb$. For a fixed $\tfix$, its first order Taylor expansion
  \begin{equation}\label{eq:T-taylor}
    \tsc_x = \tfix + \gsc (x) \delta T + \mathcal{O}(\delta T^2), \qquad x\in 
    \Lambda_N\, ,
  \end{equation}
is such that $\mathbf v= \bfgsc$ is the unique minimizer of 
$Q(\mathbf{v}; \tfix)$
for fixed $v(x)=\pm\frac{1}{2}$, $x\in \partial \Lambda_N$, where we choose
the $+$-sign for $x\in \partiallb \Lambda_N$, and $-$ for 
$x\in \partialrb \Lambda_N$.
\end{theo}

Consequently, the self-consistent profile minimizes the entropy production
up to errors of the order of $\delta T^3$.  In particular, the leading term
of the self-consistent profile can be obtained by minimization of the
entropy production. This is consistent with the general belief that for small
deviations from the equilibrium state imposed by external constraints, the
stationary state will be such that it minimizes the entropy production with
respect to variation in the unconstrained parameters \cite{KLS84}.
The entropy production has also been studied by Bodineau and Lefevere
\cite{BL08} in this model, and originally by Maes, et al.,
\cite{MNV03} in the context of heat conduction networks.

We define the thermal conductivity in the self-consistent stationary state (of
the finite system) as 
\begin{equation}
  \label{eq:fc}
    \kappa_N^{\rm sc} (\tfix)
  = \lim_{\delta T \to 0} \frac{2 N{+}1}{\delta T}\; \bar j_N .
\end{equation}
This is related to the entropy production by \eqref{eq:38}, yielding
\begin{equation}
  \label{eq:47}
  \kappa_N^{\rm sc}(\tfix) = Q(\bfgsc; \tfix)/ (2N{+}1)^{d-2}\, ,
\end{equation}
where, as in Theorem \ref{S-taylor}, we have defined
\begin{equation}
  \label{eq:48}
  Q(\mathbf v; \tfix) = \mathbf v\cdot \mathcal{J}(\tfix) \mathbf v
 = \tfix^2 \lim_{\vep\to 0}  \frac{\sigma (\mathbf \tfix + \vep \mathbf v )}{\vep^2}  \, .
\end{equation}
Since $\bfgsc$ minimizes $Q(\cdot)$, we find using \eqref{eq:46}
\begin{equation}
  \label{eq:49}
   \kappa_N^{\rm sc}(\tfix ) \le  (2 N{+}1)^{2-d}
\tfix^2 \lim_{\delta T\to 0}  \frac{\sigma (\mathbf \tlin)}{\delta T^2}  \le \tfix^{-2} C_2(\gamma;2 \tfix) \, .
\end{equation}
In particular, since the bound does not depend on $N$, this proves that the
self-consistent conductivity defined by (\ref{eq:fc}) is uniformly bounded in
$N$. It also has a Green-Kubo type of representation, as summarized in the
following theorem.
\begin{theo}\label{th:kNbound}
The self-consistent conductivity is uniformly bounded in $N$ and
satisfies
  \begin{equation}
\label{eq:1stgGK}
     \kappa_N^{\rm sc}(\tfix) = \frac {1}{\tfix^2} \int_0^\infty 
    \sum_{x\in \Lambda_N'}  (-(2 N{+}1)\nabla_{\!e_1} \gsc (x))
    \mean{j_{x,x+ \bv_1\!}(t) j_{0,\bv_1\!}(0)}_{0} \; dt 
\end{equation}
where $\mean{\cdot}_{0}$ denotes the mean
over  the initial conditions distributed according to
the equilibrium measure at the temperature $\tfix$ with the time
evolution given by the dynamics corresponding to $\mathbf\tfix$,
i.e., all the reservoirs are at temperature $\tfix$.
Here $\nabla_{\!e_1} \gsc (x)=\gsc(x+\bv_1)-\gsc(x)$ denotes a discrete
gradient. 
\end{theo}

A similar Green-Kubo formula can be obtained for the entropy production
in the stationary state of the profile $\mathbf \tlin$.  We will prove that
\begin{align} 
 \label{eq:50}
& (2 N{+}1)^{2-d}
   \tfix^2 \lim_{\delta T\to 0}  \frac{\sigma (\mathbf \tlin)}{\delta T^2} 
= 
 \lim_{\delta T\to 0}  \frac{2 N{+}1}{\delta T} 
  \frac{1}{|\Lambda'_N|} \sum_{x\in \Lambda'_N} 
  \mean{j_{x,x+ \bv_1}}_{\mu(\mathbf \tlin)} 
\nonumber \\ &\quad 
= \bigl( 1 + \frac{1}{2 N} \bigr) 
\frac {1}{\tfix^2} \int_0^\infty
  \frac{1}{|\Lambda'_N|}  \sum_{x,y\in \Lambda'_N} 
    \mean{j_{y,y+\bv_1\!}(t) j_{x,x+ \bv_1\!}(0) }_{0}\; d t\, .
\end{align}
By \eqref{eq:49}, this is always an upper bound for 
$\kappa_N^{\rm sc}(\tfix)$.  
We expect the self-consistent
profile to become linear away from the boundaries in the limit $\vep\to 0$, 
and to find $\nabla_{\!e_1}  \gsc (x) \approx -\frac{1}{2 N}$,
whenever $x_1$ is not too close to $\pm N$.  Although a proof of this property
is still missing, we conjecture accordingly that both 
$\kappa^{\rm sc}_N(\tfix)$ and the right hand side of \eqref{eq:50}
have the same limit as $N\to \infty$. 

The last result concerns the Green-Kubo representation of the
conductivity in the \emph{infinite system}. 
Consider the infinite system on $\mathbb Z^d$ with all $\gamma_x = \gamma$ 
and all thermostats at temperature $\tfix$. 
This infinite dynamics has a unique invariant measure
given by the Gibbs measure on $(\mathbb R^{2d})^{\mathbb Z^d}$ at
temperature $\tfix$, defined by the usual DLR relations.
We denote also the infinite volume Gibbs measure by $\mu_{\tfix}$. 
The existence of the dynamics of this infinite system 
 in equilibrium at any given temperature can be proven by standard 
techniques (cf.\ \cite{ot}, where a similar result is proven for an
analogous system in continuous space). A proof of the existence of the
dynamics  in dimension $2$
for a certain set of non-equilibrium initial configurations is
proven in \cite{fritz}. 
Consequently, we look at the dynamics starting from this equilibrium
distribution, and let $\mathbb E$ denote the expectation over the 
corresponding stochastic process.
\begin{theo}\label{infinite}
  There is a unique limit for
  \begin{equation}
    \label{eq:gk1}
    \frac 1{\tfix^2}\
    \lim_{\lambda \to 0}\ \sum_{x \in \mathbb Z^d} \int_0^\infty
    e^{-\lambda t} \mathbb E\left[ 
      j_{x,x+\bv_1\!}(t) j_{0,\bv_1\!}(0)\right] \; dt = 
\kappa(\tfix) \le \frac{C}{\gamma} \, ,
  \end{equation}
where $C=\mathbb E\left[ 
      (V'(q_{\bv_1\!}(0)- q_{0}(0)))^2\right]/\tfix$ is finite and depends only on $\tfix$.
\end{theo}
As we have mentioned in the introduction, the above bound for the conductivity goes to zero when $\gamma\to \infty$.

As argued earlier, we expect the self-consistent conductivity and the
Green-Kubo formula for the linear profile to have the same limit as
$N\to \infty$.  However, inspecting the definition of the latter quantity
in \eqref{eq:50} shows that this limit should be given by 
\eqref{eq:gk1}, provided the current-current correlations 
$\mean{j_{x,x+ \bv_1\!}(t) j_{y,y+\bv_1\!}(0)}_{0}$ have a
sufficiently fast uniform decay both in $t$ and in the spatial
separation $|x-y|$ (the limiting infinite system dynamics are
translation 
invariant also in the first direction, which should be employed to cancel the
sum over $y$ in \eqref{eq:50}).
Therefore, we also conjecture that 
$\kappa^{\rm sc}_N(\tfix)\to \kappa(\tfix)$, at least 
along some subsequence of $N\to \infty$.

\section{Self-consistent Profiles: Existence}
\label{sec:self-cons-prof}

The following Lemma
shows that zero temperatures cannot appear in self-consistent temperature
profiles. (We will also give a second proof
of local existence of self-consistent profiles in Section \ref{sec:temp} which
does not rely on the assumptions made about profiles containing zero
temperatures.)  
\begin{lemma}
  \label{lem:nozero}
If $\{T_x, x\in\Lambda_N\}$ are not all identically zero, then
$\left<p^2_y\right> >0$ for all $y\in\Lambda_N$.
\end{lemma}
 
\begin{proof}
  This is a consequence of the smoothness of the density of the
  transition probability $P_t(q',p'; q,p)$ of the process. Since 
  $\int P_t(q',p'; q,p)\; dq\; dp =1$, for any $(q',p')$ there exists an
  open set of positive Lebesgue measure $A = A(q',p',t)$ such that
  \begin{equation}\label{eq:tran}
    \int_A P_t(q',p'; q,p)\; dq\; dp > 0  \, .
  \end{equation}
  If there exists $x$ such that $\left<p^2_x\right> =0$, then 
  \begin{equation}
    0 = \int \mu(\mathbf T; dq', dp') \int p^2_x  P_t(q',p'; q,p)\; dq\; dp
  \end{equation}
  which clearly is in contradiction with \eqref{eq:tran}.
\end{proof}

\noindent
\emph{Proof of Theorem \ref{th:scexists}.}
Given any collection of parameters $u\in [0,\infty)^{R^c}$, $x\in R^c$,
let us define the corresponding temperature profile $\mathbf{T}(u)$ by
\begin{align} 
  T(u)_x =T(u;\Theta)_x = \begin{cases} u_x, & \text{if }x\in R^c,\\
 \Theta_x,&\text{if } x\in R .
\end{cases}
\end{align}
As before, we denote the density
of the corresponding stationary measure  by $f_N(q,p; \mathbf T(u), V, W)$.  
We have seen in the section \ref{sec:model} that,
by the hypoelliptic properties of the dynamics (cf.\ \cite{msh}),
$f_N$ is a smooth function of $(q,p)$ and consequently of
$\mathbf T$. 
By a straightforward scaling argument, we then have
for any $u$ and $\lambda>0$, 
\begin{equation}
  \label{eq:17}
  \lambda^{M} f_N(\sqrt{\lambda} q, \sqrt{\lambda} p;
  \mathbf{T}(u), V, W) 
  =  f_N(q, p; \mathbf{T}(u)/\lambda,V_\lambda, W_\lambda) 
\end{equation}
where $V_\lambda(q) = \lambda^{-1}V(\sqrt \lambda q)$ and
$W_\lambda(q) =  \lambda^{-1} W(\sqrt \lambda q)$. 
An argument similar to that used at the end of section
\ref{sec:model} to prove regularity in $\mathbf T$ shows that
$f_N(q, p; \mathbf{T}(u)/\lambda,V_\lambda, W_\lambda)$ 
is smooth in $\lambda$. Under the conditions assumed on $V$
and $W$, we have
$\lim_{\lambda\to \infty} V_\lambda(q) = V_\infty(q)$ and
$\lim_{\lambda\to \infty} W_\lambda(q) = W_\infty(q)$ with
$V_\infty(q) = \frac{1}{2}V''_\infty q^2$ and
$W_\infty(q) = \frac{1}{2} W''_\infty q^2$.

We apply the scaling relation to prove that for high enough temperatures the
system behaves essentially like a Gaussian.  More precisely, consider 
arbitrary sequences $\lambda_n\to \infty$ and 
$\mathbf{b}^{(n)}\in [0,\infty)^{\Lambda_N}$,
such that $\mathbf{b}^{(n)}$ converges to 
$\mathbf{b}\in [0,\infty)^{\Lambda_N}$.  Define further 
$T^{(n)}_x = \lambda_n b^{(n)}_x$, $x\in\Lambda_N$.  
Then  by the scaling relation (\ref{eq:17}), 
for any $x'$, 
\begin{equation}\label{eq:der17a}
  \begin{split}
    \frac{1}{\lambda_n}\left< p_{x'}^2 \right> & ( \mathbf{T}^{(n)}, V, W) 
    = \left< p_{x'}^2 \right> ( \mathbf{b}^{(n)},
    V_{\lambda_n}, W_{\lambda_n})
    \ \mathop{\longrightarrow}_{n\to\infty}\
    \left< p_{x'}^2 \right> ( \mathbf{b}, V_{\infty}, W_{\infty}) \, .
  \end{split}
\end{equation}
The last expectation is with respect to the stationary state of a purely
harmonic system.  This system was studied in \cite{bll}, where it was
proved, in Sections 3 and 7, that there is a doubly stochastic matrix $M$,
with strictly positive entries, such that for any profile of 
temperatures $\mathbf b$ and for all $x'$,
\begin{align*}
\left< p_{x'}^2 \right> ( \mathbf{b}, V_{\infty}, W_{\infty}) = 
\sum_{y\in \Lambda_N} M_{x'y} b_y.
\end{align*}
(Strictly speaking, the result was proven only for periodic profiles 
in \cite{bll}.  However, the above properties, linearity in $\mathbf b$, as
well as positivity and double stochasticity of $M$, 
are easily generalized for non-periodic profiles, although we do not go into
details here.) 
Since $\sum_y M_{xy}=1$ for all $x$, this implies 
\begin{align}\label{eq:der17b}
\left< p_{x'}^2 \right> ( \mathbf{b}, V_{\infty}, W_{\infty})
\le \max_y b_y 
 = \norm{\mathbf b}_\infty \, ,
\end{align}
and the equality holds if and only if $\mathbf b$ is a constant vector, i.e., 
$b_x$ is independent of $x$.

We can now prove the existence of a self-consistent profile.
Let $R^c=\Lambda_N \setminus R$, and consider the mapping
$F:X\to X$, $X=[0,\infty)^{R^c}$ defined
for $u\in [0,\infty)^{R^c}$, $x\in R^c$, by
\begin{align} 
  F(u)_x = \mean{p_x^2}(\mathbf{T}(u),V,W)\, .
\end{align}
Since some of the temperatures are kept fixed to non-zero values,
the hypoelliptic properties of $L^{*(1)}_N$ imply that
$F$ is everywhere continuous.
For any $L>0$
define $X_L=[0,L]^{R^c}\subset X$.  
We will soon prove that
there is an $L>0$ such that $F(X_L)\subset X_L$.  Since $X_L$ is 
homeomorphic to the unit ball of $\R^{|R^c|}$ and $F$ is continuous on
$X_L$,
we can conclude from the Brouwer fixed point theorem that there is at least one
$u\in X_L$ such that $F(u)=u$.  
By Lemma \ref{lem:nozero}, if there is $x$ such that $u_x=0$,
then $F(u)_x >0$, and such $u$ cannot be fixed points.
Thus for any fixed point $0< u_x\le L<\infty$ 
for all $x$, and $\mathbf{T}(u)$ is then a proper
self-consistent temperature profile.

We prove the existence of a constant $L$, for which $F(X_L)\subset X_L$,
by contradiction.  If no such $L$ exists, then  
for all $L>0$ there is $u^{(L)}\in X_L$ such that
$\norm{F(u^{(L)})}_{\infty}>L$.   Then necessarily
$\norm{u^{(L)}}_\infty\to \infty$, since otherwise there would
exists a convergent subsequence, which is incompatible with
$\norm{F(u^{(L)})}_{\infty}\to \infty$.  Let $\lambda_L=\norm{u^{(L)}}_\infty$
and
$v^{(L)}=\lambda_L^{-1} u^{(L)}$, so that $\lambda_L\to\infty$ and
$\norm{v^{(L)}}_\infty=1$.
The sequence $(v^{(L)})$ belongs to a compact subset of $X$, and 
we can find a subsequence such that $v^{(L)}\to v$ 
in $X$.  For this final subsequence we can apply 
(\ref{eq:der17a}) and (\ref{eq:der17b}), which shows that for all $x$
\begin{align} 
\limsup_{L} \lambda_L^{-1} F(\lambda_L v^{(L)})_x < \norm{v}_\infty=1\, .
\end{align}
Equality is not possible here,
as the limit $\mathbf{b}$ of $\lambda_L^{-1} \mathbf{T}(\lambda_L v^{(L)})$
has at least one component equal to one, but
$b_x=0$ for all $x\in R$, and thus $\mathbf{b}$
cannot be a constant vector.
However, by construction, for every $L$ there is $x(L)$ such that 
$F(\lambda_L v^{(L)})_{x(L)}>L\ge \norm{u(L)}_\infty = \lambda_L$, which leads to 
contradiction.  This proves 
the existence of $L>0$ 
with the required properties and concludes the proof of the first part of the 
theorem. 

For the second part, let us first point out that, if $R$ is invariant under
all periodic translations of $\Lambda_N$, it must be of the form 
$R=R_1\times I_N^{d-1}$, where $I_N=\set{-N,\dots,N}$ and $R_1\subset I_N$ is
non-empty.  Similarly, $\Theta_x$ can only depend on $x_1$.
Let $R_1^c=I_N\setminus R_1$, let $P_1$ denote the projection 
on the first axis in $\Z^d$, and define 
$R' = P_1 R^c = R_1^c\times\set{{\bf 0}}$, 
which is a subset of $R^c=\Lambda_N\setminus R$.  
If $R'$ is empty, $R=\Lambda_N$
and there is nothing to prove.  Otherwise, let us consider the map 
$F':X'\to X'$, $X'=[0,\infty)^{R'}$, defined by 
$F'(u)_x = \mean{p_x^2}(\mathbf{T}'(u),V,W)$, where 
\begin{align} 
  T'(u)_x = \begin{cases} u_{P_1 x}, & \text{if }x\in R^c,\\  
    \Theta_x,&\text{otherwise.}
\end{cases}
\end{align}
Every such $T'(u)$ is clearly invariant under all periodic translations.
We can then repeat the analysis made above for $F'$ and conclude that it has a
fixed point $\bar{u}$ with $0<\bar{u}_x<\infty$.
Since $\bar{T}=T'(\bar{u})$ is periodic, the dynamics is
completely 
invariant under periodic translations, implying that also expectation
values in the unique stationary state are invariant.  Therefore,
for any $x\in R^c$, we have $\mean{p_x^2}(\bar{T})=
\mean{p_{P_1 x}^2}(\bar{T})=u_{P_1 x}=\bar{T}_x$.  This proves that
$\bar{T}$  is an invariant, self-consistent profile.
\qed

\section{Entropy Production Bound}
\label{sec:entropyprod}

In this section we prove the entropy production bounds stated in Theorems
\ref{entropybound} and \ref{lin-ep}.
Given a generic profile of temperatures $\mathbf T$, we recall the
notation $f_N = f_N(\mathbf T)$ for the density of the stationary measure
$\mu({\mathbf T})$
with respect to Lebesgue measure, and let $\mean{\cdot}$ denote expectation
with respect to $\mu({\mathbf T})$.  A simple computation
shows that 
$\mean{A\ln f_N}=0$  for $A$ defined in (\ref{eq:3}).  
Therefore, by stationarity we have
\begin{equation}
  \label{eq:22}
  \begin{split}
    0 = -  \left<L_N \ln f_N\right> = 
    -  \sum_{x}  \left<S_x \ln f_N \right>
  \end{split}
 \end{equation}
where $S_x = \gamma_x(T_x \partial_{p_x}^2 - p_x  \partial_{p_x})$. Let
$\psi_x = f_N/G_{T_x}$, where $G_{T}= Z_T^{-1} \rme^{-\mathcal H_N/T}$, as 
in \eqref{eq:28}. Then we can rewrite the last term as 
\begin{equation}
  \label{eq:23}
     - \left<S_x \ln f_N \right> = 
     - \int \left( S_x \ln \psi_x\right) \psi_x  
       G_{T_x} \rmd p\, \rmd q - \int S_x ( \ln G_{T_x}  ) f_N 
       \rmd p\, \rmd q \, .
\end{equation}
Since  $p_x G_{T_x} = - T_x \partial_{p_x} G_{T_x}$ and
$S_x(\ln G_{T_x})=-\gamma_x (T_x-p_x^2)/T_x=-J_x/T_x$, we find by integration
by parts that
\begin{equation} 
  \label{eq:23b}
  \begin{split}
     - \left<S_x \ln f_N \right> 
     = T_x\gamma_x \int \frac {(\partial_{p_x}\psi_x)^2}{\psi_x}  G_{T_x}
     \rmd p\, \rmd q +  \frac{\mean{J_x}}{T_x}\, .  
  \end{split}
\end{equation}
So by (\ref{eq:22}), the entropy production satisfies
\begin{equation}
  \label{eq:ep1}
 \sigma (\mathbf T) =  -\sum_{x\in \Lambda_N} 
 \frac{\mean{J_x}}{T_x} =
  \sum_{x\in \Lambda_N} 
\mathcal D_x \, ,
\end{equation}
where
\begin{equation}
  \label{eq:defDx}
\mathcal D_x = \gamma_x  T_x \int \frac {(\partial_{p_x}\psi_x)^2}{\psi_x}
G_{T_x}  \rmd p\, \rmd q\, .
\end{equation}
In particular, $\sigma (\mathbf T)\ge 0$, and by 
using the local conservation of energy, \eqref{eq:5}, 
\eqref{eq:29a} holds.

Let us for the remainder of this section assume that $\mathbf T$ is
a temperature profile which is invariant under the periodic translations.
The results then hold for both $\mathbf\tsc$ and $\mathbf\tlin$.
Obviously, then by \eqref{eq:29a}
\begin{equation}
  \label{eq:29}
  \sigma (\mathbf T) =  
  \sum_{x\in \Lambda'_N} (\beta_{x+e_1} - \beta_{x}) \mean{j_{x,x+e_1}} 
\end{equation}
Therefore, it will suffice to find a bound for $|\mean{j_{x,x+e_1}}|$.

Applying the definition of the current observable, (\ref{eq:6})
and (\ref{eq:6b}), and then
integration by parts, shows that
  \begin{align}  \label{eq:30}
 &   \left<j_{x,x+\bv_1}\right> = - \frac{1}{2} 
\int V'(r_x) \sum_{n=0}^1 \left. \psi_{x'} p_{x'} G_{T_{x'}} 
\right|_{x'=x+n e_1}  \rmd p\, \rmd q \nonumber \\
& \quad = -\sum_{n=0}^1  \frac{T_{x'}}{2} 
\int V'(r_x) G_{T_{x'}}  \partial_{p_{x'}}\psi_{x'}
  \rmd p\, \rmd q  \Bigr|_{x'=x+n e_1}
  \end{align} 
where $r_x = q_{x+\bv_1}-q_x$.  We use that
$1=\psi_{x'}^{1/2}/\psi_{x'}^{1/2}$ whenever $\psi_{x'}\ne 0$,
and then apply the Schwarz inequality.  This shows that 
  \begin{align} 
  \label{eq:36}
 | \mean{j_{x,x+\bv_1}} |^2 \le 
 \max_{y\in{\Lambda_N}}\frac{T_y}{\gamma_y}
  \left< V'(r_x)^2 \right>
  \frac{1}{2} \sum_{n=0}^1  \mathcal D_{x+n e_1} \, .
  \end{align}
Therefore, we have obtained the following relation between the 
total sum of currents and the entropy production
\begin{align} 
  \label{eq:37}
 &  \Bigl(\sum_{x\in \Lambda'_N} | \mean{j_{x,x+\bv_1}} | \Bigr)^2
 \le  
 \max_{y\in{\Lambda_N}}\frac{T_y}{\gamma_y}
 \sum_{x\in \Lambda'_N} \left< V'(r_x)^2 \right>
 \sum_{x\in \Lambda'_N}
   \frac{1}{2} \sum_{n=0}^1  \mathcal D_{x+n e_1} \nonumber\\
& \quad \le
\sigma (\mathbf T) \max_{y\in{\Lambda_N}}\frac{T_y}{\gamma_y}
 \sum_{x\in \Lambda'_N} \left< V'(r_x)^2 \right>
  \,.
\end{align}

For this bound to be useful, we still need to consider 
$\sum_{x\in \Lambda'_N} \left< V'(r_x)^2 \right>$.
Since $L_N(q_x^2) = 2 q_x p_x$, we have $\mean{q_x p_x}=0$ for all $x$.
Similarly, $L_N \mathcal{H} = \sum_{x\in \Lambda_N} \gamma_x (T_x - p_x^2)$
implies $\sum_{x} \gamma_x T_x = \sum_{x} \gamma_x \mean{p_x^2}$. Now
\begin{align} 
  L_N(\sum_{x\in \Lambda_N}  p_x q_x) = 
  \sum_{x\in \Lambda_N} p_x^2
  - \sum_{x\in \Lambda_N} q_x \partial_{q_x} \mathcal{H}
 - \sum_{x\in \Lambda_N} \gamma_x p_x q_x \,,
\end{align}
and thus
\begin{align}\label{eq:gTx}
\sum_{x\in \Lambda_N} \gamma_x T_x
 \ge \min_y \gamma_y \sum_{x\in \Lambda_N} \mean{p_x^2}
 = \min_y \gamma_y 
\left< \sum_{x\in \Lambda_N} q_x \partial_{q_x} \mathcal{H} \right> \, .
\end{align}
From the asymptotics of $V$ and $W$ we can conclude that there are
$C>0$ and $C'\ge 0$ such that
\begin{align} 
  V'(r)^2 \le C ( r V'(r) + C' ) \quad\text{and}\quad
 r W'(r) \ge -C'  \,.
\end{align}
But since
\begin{align}\label{eq:pqxH}
 &  \partial_{q_x} \mathcal{H} = W'(q_x) +
 \sum_{j=1}^d \left(V'(q_x-q_{x-\bv_j})-V'(q_{x+\bv_j}-q_x)\right) 
\nonumber \\
&\qquad 
 +\frac{1}{2}\left(\mathbbm{1}(x\in \partialrb \Lambda_N)
   V'(-q_x)-\mathbbm{1}(x\in \partiallb \Lambda_N)
   V'(q_x)
 \right) \, ,
\end{align}
with $\mathbbm{1}$ denoting the characteristic function, we have
\begin{align} 
&\sum_{x\in \Lambda_N} q_x \partial_{q_x} \mathcal{H}
= \sum_{x\in \Lambda_N} \Bigl[ q_x W' (q_x)
  + \sum_{j=2}^d \left. r V'(r)\right|_{r=q_{x+e_j-q_x}} \Bigr] 
\nonumber \\
&\qquad 
+ \sum_{x\in \Lambda'_N} r_x V'(r_x)
+\frac{1}{2} \sum_{x\in \partialrb\Lambda} q_x V'(q_x)
+\frac{1}{2} \sum_{x\in \partiallb\Lambda} (-q_x) V'(-q_x) \, .
\nonumber \\
&\quad 
\ge \sum_{x\in \Lambda'_N} r_x V'(r_x)
 - |\Lambda_N| C' (d+1)\, .
\end{align}
Combining this with (\ref{eq:gTx}) shows that
\begin{align} 
 \sum_{x\in\Lambda'_N} 
  \left< V'(r_x)^2 \right> \le C |\Lambda_N|
  \left( C' (d+2) + \frac{\max_y \gamma_y T_y}{\min_y\gamma_y}\right) \,.
\end{align}
Consequently, there is $c>0$, which depends only on $V$ and $W$,
such that
\begin{align} 
  \label{eq:37b}
 &  \Bigl(\sum_{x\in \Lambda'_N} | \mean{j_{x,x+\bv_1}} | \Bigr)^2
 \le c \sigma (\mathbf T) |\Lambda_N|
   \frac{\max_x \gamma_x T_x}{\min_x \gamma_x^2} ( 1 + \max_x T_x) \, .
\end{align}

Let us next consider the case $\mathbf T=\mathbf \tlin$.
Applying the definition of $\mathbf \tlin$ to \eqref{eq:29} shows that then
\eqref{eq:45} holds, i.e., 
$\sigma (\mathbf \tlin ) =  \frac{\betar -\betal}{2N} 
   \sum_{x\in \Lambda'_N} \mean{j_{x,x+e_i}}$.
Then by \eqref{eq:37b} and using the fact that $\tlin_x \le \tlb$
\begin{equation}
  \label{eq:41}
  \sigma (\mathbf \tlin )\le c'
 \left|\betar-\betal\right|^2  (2 N+1)^{d-2} (1+\tlb)^2  \, ,
\end{equation}
where $c'$ is a constant depending only on $\gamma$, $V$, and $W$.
Therefore, we have now proven Theorem \ref{lin-ep}.

Finally, let us consider the self-consistent case, 
$\mathbf T=\mathbf \tsc$.  
For the corresponding stationary measure we find from (\ref{eq:5}),
 \begin{equation}
   \label{eq:33}
   \sum_{j=1}^d  \left( \left<j_{x, x+\bv_j}\right> - \left<j_{x-\bv_j,
       x}\right>\right) = 0, \quad x\not\in\partial\Lambda_N \, .
 \end{equation}
Since the system, including the self-consistent profile, 
is periodic in any of the Cartesian directions $\bv_i$, $i\ge 2$, 
also the unique stationary measures are invariant under translations in these
directions.  Therefore,
\begin{equation} \label{eq:34}
   \left<j_{x, x+\bv_i}\right> = \left< j_{x-\bv_i, x}\right>, \qquad
   i\neq 1,\ x\in \Lambda_N\, .
\end{equation}
Consequently, by (\ref{eq:33}) and (\ref{eq:5}),
 \begin{equation}
   \label{eq:666}
   \begin{split}
     \left<j_{x, x+\bv_1}\right> & = \left<j_{x-\bv_1, x}\right>,\qquad 
     x\not \in\partial\Lambda_N\, ,  \\
     \left<j_{x, x+\bv_1}\right> & = \mean{J_x} =
\gamma_x(\tlb - \left<p_x^2\right>),  \qquad 
     x \in \partiallb \Lambda_N\, , \\
     \left<j_{x- \bv_1, x}\right> &= -\mean{J_x} =
\gamma_x(\left<p_x^2\right> - \trb), \qquad 
     x \in \partialrb \Lambda_N \, .
   \end{split}
 \end{equation}
We denote the constant current by $\bar{j}_N$, i.e., now we have
$\mean{j_{x,x+\bv_1}} = \bar{j}_N$, for all $x\in \Lambda'_N$.
Therefore, by (\ref{eq:29}),
\begin{align}  
  \label{eq:244}
 &  \sigma (\mathbf \tsc) = \bar{j}_N 
  \sum_{x\in \Lambda'_N} (\beta_{x+e_1} - \beta_{x})
= \bar{j}_N (\betar-\betal) (2 N +1 )^{d-1}\, ,
\end{align}
which proves \eqref{eq:38}.
This immediately implies that $\mathrm{sign}(\tlb-\trb) \bar{j}_N\ge 0$.
But on the other hand, 
$\bar{j}_N = \frac{1}{|\Lambda'_N|} \sum_{x\in \Lambda'_N}
\mean{j_{x,x+\bv_1}}$, and thus also for the self-consistent profile
$\sigma (\tscprof ) =  \frac{\betar -\betal}{2N} 
   \sum_{x\in \Lambda'_N} \mean{j_{x,x+e_i}}$.  Applying \eqref{eq:37b}
then completes the proof of Theorem \ref{entropybound}.

\section{Minimization of entropy production}
\label{sec:temp}

For a given $\tfix>0$, we  
use the Gibbs measure $\mu_{\tfix} = G_{\tfix} dp dq$ as a reference
measure and we denote  
the related  expectation by $\left<\cdot\right>_0$.
We consider the generator $L$ on 
the Hilbert space $L^2(\mu_{\tfix})$.  Recall that for any temperature profile
$\mathbf T = \{T_x, x\in \Lambda_N\}$ we have $L = L(\mathbf T) =A+
S(\mathbf T)$.  Its adjoint is
\begin{align} 
  L^* = -A+ \sum_{x\in \Lambda_N} S_x^*
\end{align}
where $S_x = \gamma_x\left( T_x \partial_{p_x}^2 -
  p_x \partial_{p_x}\right)$, 
and thus
\begin{align} 
 S_x^* = S_x + \gamma_x \frac{\Delta T_x}{\tfix} \left(
   - 2 p_x \partial_{p_x} + h_x \right)
\end{align}
with $\Delta T_x = T_x-\tfix$ and
\begin{align} 
 h_x = \frac{p_x^2}{\tfix} -1.
\end{align}
Observe that $\mean{h_x h_{x'}}_0 = {2} \delta_{x, x'}$ and
$-S_{x,\tfix} h_x = 2 \gamma_x h_x$.

Set $L_0 = L(\mathbf \tfix)$ and consequently
$L_0^* = -A + S(\mathbf \tfix)^* = -A + S(\mathbf \tfix)$.

\begin{lemma}
For all $y,x$
\begin{align}\label{eq:px2diff}
\left. \partial_{T_{y}} \mean{p_x^2}_{\mu(\mathbf T)} 
  \right|_{\mathbf T=\mathbf\tfix}
 = \gamma_{y} \mean{h_{y} (-L_0)^{-1} h_x}_0 
 = \gamma_{y} \mean{h_{x} (-L_0)^{-1} h_y}_0 
\, .
\end{align}
\end{lemma}
\begin{proof}
Let us denote by $f = f(\mathbf T)$ the density of $\mu(\mathbf T)$
with respect to $\mu_{\tfix}$.
Then $f$ is solution of the equation $L^*(\mathbf T) f(\mathbf T) =
0$. Since the 
coefficients in  $L^*(\mathbf T)$ are smooth in $\mathbf T$, $f$ is
smooth in $\mathbf T$ and 
$f_{y} = \partial_{T_{y}} f(\mathbf T)$ solves the equation 
\begin{align} 
L^*(\mathbf T) f_{y}(\mathbf T)
=- (\partial_{T_{y}} L^*) (\mathbf T) f(\mathbf T) = - \frac
{\gamma_{y}}{\tfix}\left( \tfix \partial_{p_{y}}^2 -2
  p_{y} \partial_{p_{y}} +  h_{y}\right) f(\mathbf T)\, .
\end{align}
Since $f(\mathbf \tfix)= 1$, we have found 
that $f_{y}(\mathbf \tfix)$ is solution of 
\begin{equation}
 - L_0^*f_{y}(\mathbf\tfix) =  \frac {\gamma_{y}}{\tfix} h_{y}\
 .\label{eq:9}  
\end{equation}
Notice that $f_{y}$ has a bounded $L^2(\mu_{\tfix})$ norm (cf.\
\cite{villani}), 
and by a standard argument (multiply equation (\ref{eq:9}) by $f_y$ and
integrate with respect to $\mu_{\tfix}$) we obtain a bound
\begin{equation}
  \sum_x \gamma_x \mean{(\partial_{p_x} f_{y})^2}_0 \le \gamma_y \tfix^{-1} .
\end{equation}
Now, since
$h_x G_{\tfix} = -\partial_{p_x}(p_x G_{\tfix})$, 
\begin{equation}
  \mean{h_x}_{\mu(\mathbf T)} = 
  \mean{ p_x \partial_{p_x} f(\mathbf T)}_0 \, .
\end{equation}
Then differentiating with respect to $T_{y}$ we have
\begin{equation}
  \label{eq:dtx}
   \partial_{T_{y}} \mean{h_x}_{\mu(\mathbf T)} = 
   \mean{p_x \partial_{p_x} f_{y}(\mathbf T)}_0 
\end{equation}
and taking the limit $\mathbf T\to \mathbf \tfix$
\begin{equation}
  \label{eq:8}
   \left.  \partial_{T_{y}} \mean{h_x}_{\mu(\mathbf
       T)}\right|_{\mathbf T=\mathbf \tfix} =
 \mean{ p_x \partial_{p_x}
     f_{y}(\mathbf\tfix)}_0 = \mean{ h_x 
     f_{y}(\mathbf\tfix)}_0 = \frac{\gamma_{y}}{\tfix} \mean{h_{x}
     (-L^*_0)^{-1} h_{y}}_0 
\end{equation}
Observe that, since $h$ is an even function of $p$, one can, by a change of
variables $p \to -p$, replace $L_0^*$ with $L_0$ in
(\ref{eq:8}). This proves \eqref{eq:px2diff}. \end{proof}

Define $F:\R_+^{\Lambda_N} \to \R^{\Lambda_N}$ as
\begin{align} 
  F_x(\mathbf T) = \mean{J_x}_{\mu(\mathbf T)} =
\gamma_x\left(T_x - \mean{p_x^2}_{\mu(\mathbf T)}\right) \, .
\end{align}
Its Jacobian at $\mathbf T = \mathbf \tfix$ is given by
\begin{equation}
  \label{eq:jaco}
   \mathcal{J}_{y,x} = \gamma_x\delta_{y,x} - \gamma_x \left. \partial_{T_{y}}
    \mean{p_x^2}_{\mu(\mathbf{T})} \right|_{\mathbf{T}=\tfix}    = \gamma_x
  \delta_{y,x} -   \gamma_x\gamma_{y} \mean{h_{x} (-L_0)^{-1} h_{y}}_0. 
\end{equation}
Observe that $\mathcal{J}$ is symmetric and that $F(\mathbf \tfix) = 0$ for any
value of $\tfix$. It follows that $0$ is an eigenvalue of $\mathcal J$, 
and we will show shortly that $\mathcal J\ge 0$, and the 
eigenspace corresponding to $0$
is one-dimensional and generated by the constant vector.
Then the matrix $M=(\mathcal{J}_{x,y})_{x,y\in R^c}$ is invertible, and thus
there is a 
neighborhood in $(\tlb,\trb)$ containing $(\tfix,\tfix)$ such that the
implicit function theorem can be applied to obtain a self-consistent
profile.  
This implies that constants $\vep_0$ and $\delta$ for the first part of
Theorem \ref{th:scexists2} can be found.  It also follows that 
$\mathbf \tsc(\tlb,\trb)$ is smooth.  To see that it must also be invariant
under the periodic translations, we first point out that in the boundary
layer setup clearly any translate of a self-consistent profile is also
self-consistent.  Since the translations correspond to a permutation of
indices, they remain in the neighborhood determined by $\delta$, and thus
by the uniqueness of the self-consistent profile in this neighborhood,
$\tsc(\tlb,\trb)$ must itself be invariant. 

Therefore, to complete the proof of Theorem \ref{th:scexists2}
we only need to prove the following Lemma.
\begin{lemma}
$\mathcal J\ge 0$, and $\mathcal{J} a=0$ implies $a_x$ is a constant in $x$.
\end{lemma}
\begin{proof}
Let $a\in \mathbb R^{\Lambda_N}$, and define 
$h = \sum_{x\in \Lambda_N} a_x h_x$. It follows from  the antisymmetry of $A$
and the symmetry of $S_0$:
\begin{equation}
  \label{eq:algebra}
  \begin{split}
    \mean{(Ah) (- L_0)^{-1} (Ah)}_0 = 
    \mean{h (-S_0) h}_0 - \mean{(S_0 h) ( - L_0)^{-1} (S_0 h)}_0 \, .
   \end{split}
\end{equation}
Since $S_0 h = -2 \sum_x a_x \gamma_x h_x$, we obtain
\begin{align} 
  \label{eq:limi}
  & \mean{(Ah) (- L_0)^{-1} (Ah)}_0 =
    \mean{h (-S_0) h}_0 - 4 \sum_{x, y} a_x a_{y}\gamma_x \gamma_{y} \mean{ h_x
      (- L_0)^{-1} h_{y}}_0\nonumber \\ & \quad
    =  4 \sum_x \gamma_x a_x^2 - 4\sum_{x, y} a_x  a_{y}\gamma_x \gamma_{y}
    \mean{ h_x (- L_0)^{-1} h_{y}}_0 
\nonumber \\ & \quad
    = 4\sum_{x, y}  a_x a_{y}  \mathcal{J}_{x, y} \ .
\end{align}
Therefore, to prove that $\mathcal J$ has the properties stated above, it
suffices to study the left hand side of (\ref{eq:algebra}), and to prove that
it is always positive, and equal to
zero if and only if $a$ is a constant vector.  (Studying real
vectors $a$ suffices here, as $\mathcal J$ is a symmetric matrix.)

In fact, define $u =  (- L_0)^{-1} (Ah)$. Since for any 
observable $F$ belonging to the domain of $A$, $\mean{F(AF)}_0=0$,
we have then
\begin{equation}
  \label{eq:10}
\mean{(Ah) (- L_0)^{-1} (Ah)}_0 = \mean{ u (-S_0) u}_0  
  = \sum_x\gamma_x \tfix \mean{ (\partial_{p_x} u)^2}_0 
\ge 0\, .
\end{equation}
This proves the required positivity.
In addition, if the left hand side is zero, then
$u(p,q)$ cannot depend on $p$, and thus
\begin{equation}
  -L_0 u= -A u= - \sum_x p_x \partial_{q_x} u(q) = Ah = - \frac 2{T_0}
\sum_{x} a_x p_x \partial_{q_x} \mathcal H\label{eq:11}
\end{equation}
It follows, for all $x$,
\begin{align} 
\frac{2}{\tfix}  a_x \partial_{q_x} \mathcal H
= \partial_{q_x} u(q) \, .
\end{align}
Thus the function
\begin{align} 
  \mathcal G(q) = \frac{\tfix}{2} u(q) - \sum_{x\in\Lambda_N} a_x W(q_x)
\end{align}
satisfies, by (\ref{eq:pqxH}),
\begin{align}\label{eq:bVpeq}
\partial_{q_x} \mathcal G(q) 
& =
 a_x \Bigl[
 \sum_{j=1}^d \left(V'(q_x-q_{x-\bv_j})-V'(q_{x+\bv_j}-q_x)\right) 
\nonumber \\ &\qquad 
 +\frac{1}{2}\mathbbm{1}(x\in \partialrb \Lambda_N)
   V'(-q_x)-\frac{1}{2}
\mathbbm{1}(x\in \partiallb \Lambda_N)   V'(q_x) \Bigr]\, .
\end{align}
For $x\in\Lambda'_N$ and $k=1,2,\ldots$ we differentiate
(\ref{eq:bVpeq}) with respect to $q_{x+\bv_k}$ and obtain
\begin{equation}
  \label{eq:12}
  -a_x V''(q_{x+\bv_k}-q_x) = \partial^2_{q_x,q_{x+\bv_k}}\mathcal G(q) =
- a_{x+\bv_k}  V''(q_{x+\bv_k}-q_x) \,.
\end{equation}
Since there exists an $r_0$ such that $V''(r_0) >0$, this implies
$a=\text{const}$. 
\end{proof}

We can now conclude that for any $\tfix>0$, there is $\vep_0>0$ such that
for all 
$|\vep|<\vep_0$ a self-consistent profile corresponding to 
$\tlb=\tfix +\frac{\vep}{2}$, $\trb=\tfix -\frac{\vep}{2}$ can be found.
This profile is differentiable with respect to $\vep$ and the derivative
satisfies for $x\not\in \partial\Lambda_N$
\begin{align} 
 0  = \frac{\partial}{\partial\vep}F_x(\mathbf{T}(\vep;\tfix)) 
= \sum_{y\in \Lambda_N} \frac{\partial T_y}{\partial\vep}
 \partial_{T_y}F_x(\mathbf{T}(\vep;\tfix)) \,.
\end{align}
Therefore, we have
$\sum_{y\in \Lambda_N} \mathcal{J}_{x,y} \frac{\partial T_y(0)}{\partial\vep}
= 0 $.  
This shows that for $x \not\in \partial \Lambda_N$,
\begin{align}\label{eq:Tsc1stpert}
 \left. \frac{\partial T_x(\vep;\tfix)}{\partial\vep}\right|_{\vep=0}
 = \sum_{y\not\in \partial \Lambda_N} (M^{-1})_{x,y}\frac{1}{2}
 \biggl(
 \sum_{y'\in \partialrb \Lambda_N} \mathcal{J}_{y,y'}
 -  \sum_{y'\in \partiallb \Lambda_N} \mathcal{J}_{y,y'} 
\biggr)\, ,
\end{align}
where $M = (\mathcal{J}_{x,y})_{x,y\not\in \partial \Lambda_N}$ is a strictly
positive 
matrix, and thus invertible.

Recall the definition of entropy production given in (\ref{eq:21}).
By (\ref{eq:ep1}) we have then always $\sigma (\mathbf T)\ge 0$,
with equality when $\mathbf T=\mathbf \tfix$, a constant profile
given by $\tfix>0$. Since
\begin{equation}
    \frac{\partial \sigma }{\partial T_x}({\mathbf T}) = 
    -\gamma_x\frac{\left<p_x^2\right>}{T_x^2} + \sum_y \gamma_y
    \frac{ \partial_{T_x}\left<p_y^2\right>}{T_y} \, ,
\end{equation}
we have for the constant profile
\begin{align} 
  &  \frac{\partial \sigma }{\partial T_x}({\mathbf \tfix}) = - \tfix^{-1}
    \gamma_x + \tfix^{-1} \frac{\partial}{\partial T_x}
    \biggl(\sum_y \gamma_y \left<p_y^2\right>
    \biggr)_{\mathbf T = \mathbf \tfix}  
 \,.
\end{align}
As mentioned earlier, for any profile 
$\sum_y \gamma_y \left<p_y^2\right> = \sum_y \gamma_y T_y$,
and thus we have proven that
\begin{equation}
  \label{eq:26}
  \frac{\partial \sigma }{\partial T_x}(\mathbf \tfix) =  0 \, .
\end{equation}
A similar, but a slightly longer calculation, shows that 
\begin{equation}
  \label{eq:27}
   \frac{\partial^2 \sigma }{\partial T_x\partial T_y }( 
   {\mathbf\tfix}) = \frac{1}{\tfix^2} (
   \mathcal{J}_{x,y}+\mathcal {J}_{y,x})=
   \frac{2}{\tfix^2} \mathcal{J}_{x,y} \,. 
\end{equation}

By dividing $\Lambda_N$ into $R\ne \emptyset$ (the fixed thermostats) and
$R^c$, we can conclude from the previous results 
that the symmetric matrix $M=(\mathcal{J}_{x,y})_{x,y \in R^c}$ is strictly
positive.   By (\ref{eq:26})  and (\ref{eq:27}),
the Taylor expansion
of $\sigma $ around $\mathbf \tfix$ yields
\begin{align}\label{eq:Staylor2}
\sigma (\mathbf \tfix + \vep\mathbf{v}) 
= \frac{\vep^2}{\tfix^2} 
\sum_{x,y\in \Lambda_N} \mathcal{J}_{x,y} v_x v_y + \mathcal{O}(\vep^3)\, .
\end{align}
This proves Theorem \ref{S-taylor}.
For fixed $\vep$ and $v_x$, $x\in R$, 
the quadratic form corresponding to the leading term has a unique minimizer,
given by $\vmin_x=v_x$, $x\in R$, and
\begin{align}\label{eq:vmindef}
    \vmin_x = -\sum_{y\in R^c} \sum_{y'\in R} (M^{-1})_{xy} 
      \mathcal{J}_{y,y'} v_{y'}\, ,
 \quad\text{for }x\in R^{c} \, .
\end{align}

Let us next consider the case studied earlier, with the opposite boundaries
fixed at two different temperatures $\tlb$ and $\trb$.  Denote
$\delta T = \tlb - \trb$, which we assume to be positive, and
$\tfix = (\tlb + \trb)/2$.  Let us consider a sequence of $\tlb,\trb$ for which
$\tfix$ remains fixed and $\delta T\to 0$.  We assume that
$\mathbf T$ is a sequence of profiles with boundary values on $R$ 
equal to $\tlb$ and $\trb$, and which has a Taylor expansion
  \begin{align}\label{eq:TminS}
  T_x = \tfix + g (x) \delta T + \mathcal{O}(\delta T^2)
  \end{align}
where $g$ is a function for which $g (x) = 1/2$ for 
$x\in\partiallb\Lambda_N$ 
and 
$g (x) = -1/2$ for $x\in  \partialrb\Lambda_N$.
By (\ref{eq:Staylor2}), the entropy production will be of the order 
$(\delta T)^2$, 
and the leading term is minimized by $g (x)=\vmin_x$ corresponding to
$v_x=\pm \frac{1}{2}$,
with $+$, if $x\in \partiallb \Lambda_N$,
and $-$, if $x\in \partialrb \Lambda_N$.

We have proven in the beginning of this section, that
the self-consistent profile can be chosen for all sufficiently small 
$\delta T$ so that it is differentiable in the boundary temperatures.
In particular, comparing (\ref{eq:Tsc1stpert}) to (\ref{eq:vmindef}) shows that
\begin{equation}
   T^{\text{sc}}_x = \tfix + \gsc (x) \delta T + \mathcal{O}(\delta T^2)
\end{equation}
with $\bfgsc=\mathbf \vmin$.  We have thus proven Theorem \ref{T-taylor}.

\section{Conductivity of the Finite System}
\label{sec:cond}

In the following we again set
$\Lambda'_N =\Lambda_N\backslash \partialrb\Lambda_N$, and consider, as in
Section \ref{sec:entropyprod}, 
a generic profile $\mathbf T$ which is invariant under periodic translations.
Let $\mean{\cdot}$ be the expectation with respect to the
corresponding stationary state.
It is convenient now to use as a reference measure the
inhomogeneous Gibbs measure 
$\nu_{\mathbf T} =G(\mathbf T;q,p) \rmd q\rmd p$, with
  \begin{equation}
    G(\mathbf T;q,p) = \frac{\exp(-\sum_x \en_x(q,p)/T_x )}{Z} 
  \end{equation}
where $\en_x$ is defined in (\ref{eq:1}).
Notice that $S$ is automatically symmetric with respect to 
$\nu_{\mathbf T}$, while the adjoint of $A$ is given by
  \begin{equation}
    -A + \sum_{x\in \Lambda'_N} \Bigl( \frac{1}{T_{ x+ e_1}}-\frac{1}{T_{x}}
    \Bigr)
    j_{x,x+e_1} \, .
  \end{equation}
 
Let us next inspect $\mathbf T = \tscprof$ and 
denote by $\tilde{f}$ the density of the self-consistent stationary
state with respect to $\nu_{\tscprof}$. Let us fix $\tfix=\frac{\trb+\tlb}{2}$
with $\vep=\delta T = \tlb-\trb>0$, as before. Repeating the argument used in
section \ref{sec:model}, we find that
$\tilde{f}$ is smooth in $\vep$, so a first order development
in $\vep$ is justified.  Using the expansion of the self-consistent profile,
\eqref{eq:TminS}, shows that  
$u = \partial_{\vep} \tilde{f} |_{\vep=0}$ is solution of the equation
 \begin{equation}\label{adj1}
   (-A + S(\mathbf\tfix) ) u = 
   \sum_{x\in \Lambda'_N} \frac{\nabla_{\!e_1} \gsc (x)}{\tfix^2}
 j_{x,x+e_1} \, .
 \end{equation}
Explicit formulae for the derivatives of the self-consistent profile,
$\gsc (x)$, are given in (\ref{eq:Tsc1stpert}). 

Recall the definition of
the conductivity of the finite system, (\ref{eq:fc}).  
Since we have already proven Theorems 1--\ref{T-taylor}, the argument given
before Theorem \ref {th:kNbound} in Section \ref{sec:summary-results}
provides a proof that $\kappa_N(\tfix)$ is bounded in $N$.
On the other hand, by \eqref{adj1},
  \begin{align} 
  \label{eq:fc1}
   &  \kappa_N(\tfix) = \lim_{\delta T \to 0} \frac{2 N+1}{\delta T} 
    \frac{1}{|\Lambda'_N|} \sum_{x\in \Lambda'_N} 
    \left<j_{x,x+\bv_1}\right> 
\nonumber \\ &\quad 
= \lim_{\delta T \to 0} \frac{2 N+1}{\delta T} 
  \left<j_{0,\bv_1\!}\right>
     = (2 N+1) \mean{u j_{0,\bv_1}}_0 \, .
  \end{align}
Define $\check u (q,p) = u(q,-p)$, and observe that, since $j_{x,x+\bv_1}$
is antisymmetric in $p$,
\begin{align} 
  (A + S(\mathbf\tfix) ) \check u = 
   -\sum_{x\in \Lambda'_N} \frac{\nabla_{\!e_1} \gsc (x)}{\tfix^2}
 j_{x,x+e_1} \, .
\end{align}
Thus
\begin{equation}
  \begin{split}
    \kappa_N(\tfix) &= - (2 N+1) \mean{\check u j_{0,\bv_1}}_0 
  =  (2 N+1) \int_0^\infty \partial_t
  \mean{\check u(t) j_{0,\bv_1}(0)}_0\; dt
    \\
    &= \frac
    {1}{\tfix^2} \int_0^\infty \sum_{x\in \Lambda_N'} 
    (-(2 N+1)\nabla_{\!e_1} \gsc (x)) \mean{j_{x,x+ \bv_1}(t)
      j_{0,\bv_1\!}(0)}_{0} \; dt
  \end{split}
\end{equation}
where $\mean{\cdot}_{0}$ denotes taking the initial data distribution
according to 
the equilibrium measure at the specified temperature $\tfix$,
and then considering the time-evolution corresponding to the stochastic
process with all heat-bath temperatures set to $\tfix$.
We have used here the property that then
$\mean{\check u(t) j_{0,\bv_1}(0)}_0\to 
\mean{\check u}_0 \mean{j_{0,\bv_1}}_0=0$ for $t\to\infty$.
This completes the proof of Theorem \ref{th:kNbound}.

Repeating the same steps for $\mathbf T=\mathbf \tlin$, for which 
$\partial_\vep  \tlin_x\bigr|_{\vep=0} = -\frac{x_1}{2 N}$,
proves also the validity of (\ref{eq:50}).

\section{Conductivity of the Infinite System}
\label{sec:cond-infin-syst}

We prove here Theorem \ref{infinite} concerning the infinite system on 
$(\mathbb R^{2d})^{\mathbb Z^d}$ with all $\gamma_x = \gamma$ 
and all thermostats at temperature $\tfix$. 
This infinite dynamics has a unique invariant measure
given by the Gibbs measure on $(\mathbb R^{2d})^{\mathbb Z^d}$ at
temperature $\tfix$, defined by the usual DLR relations. We denote this measure by $\mu_{\tfix}$ and its expectation by $\left<\cdot\right>_0$.
Consequently we look at the dynamics starting from this equilibrium
distribution.

We adapt here an argument used in \cite{ben}.
Introduce on $L^2(\mu_{\tfix})$ a degenerate scalar product
\begin{equation}
  \label{eq:sp}
  \dmean{\varphi,\psi} = 
  \sum_{x\in \Z^d} [\left<\varphi \tau_x\psi\right>_0 - \left<\varphi\right>_0
  \left<\psi\right>_0]\, ,
\end{equation}
where $\tau_x$ is the translation operator.  The scalar product can also be
obtained via the limit 
\begin{equation}
  \label{eq:4}
  \dmean{\varphi,\psi} = \lim_{n \to \infty}
  {\rm Cov}_{\mu_{\tfix}}(\Phi_n \varphi,\Phi_n\psi)
= \lim_{n \to \infty} \left(\mean{\Phi_n \varphi \,\Phi_n\psi}_0
 -\mean{\Phi_n \varphi }_0\mean{\Phi_n\psi}_0\right)\, ,
\end{equation}
where  $\Phi_n$ maps functions into the corresponding ``fluctuation averages''
in $\Lambda_n$, a square box of linear size $n$ centered at $0$.  Explicitly,
\begin{align}
 (\Phi_n \psi)(q,p)=\frac {1}{\sqrt{|\Lambda_n|}} 
    \sum_{x\in\Lambda_n} (\tau_x \psi)(q,p)\, .
\end{align}
The scalar product is degenerate, since
every function 
of the form $\phi = \psi - \tau_x\psi$ is in its kernel.
We denote by $\mathcal L^2$ the corresponding Hilbert space of square
integrable functions. More precisely, $\mathcal L^2$ is a space of
classes of functions such that each of its elements can be identified with 
a function in  $L^2(\mu_{\tfix})$
up to a translation.

 Observe that $A$ and $S$ are still
respectively anti-symmetric and symmetric with respect to the scalar
product $\dmean{\cdot,\cdot} $. We also introduce the semi-norm
\begin{equation}
  \label{eq:h1norm}
  \| \varphi \|_1^2 = \dmean{ \varphi, (-S) \varphi } 
\end{equation}
and let $\mathcal H_1$ denote the corresponding Hilbert space obtained
by closing $\mathcal L^2$ with respect to  $\| \cdot \|_1$.  To see that $\|
\varphi \|_1$ is a semi-norm,
in particular, that it is positive, we can employ the easily derived identity 
\begin{equation}
  \label{eq:7}
   \| \varphi \|_1^2 = \lim_{n \to \infty} \left< \Phi_n \varphi \, (-S  \Phi_n \varphi) \right>_0  \, .
\end{equation}
Since $S$ acts only on velocities, $ \| \cdot \|_1$ has a kernel
consisting of  all functions which depend only on $q$,
the position variables. Thus also $\mathcal H_1$ is 
a space of equivalence classes of functions.   

Let $\lambda>0$ be given and let 
$u_\lambda$ be the solution of the resolvent equation
\begin{equation}
  \label{eq:res}
  \lambda u_\lambda - L u_\lambda = j_{0,\bv_1} \, .
\end{equation}
The solution can be given explicitly in terms of the 
the semigroup  $P^t$ generated by $L= A+ S$, 
\begin{equation}
  \label{eq:13}
  u_\lambda(q,p) = \int_0^\infty e^{-\lambda t} (P^t j_{0,\bv_1}) (q,p) \; dt\, .
\end{equation}
Obviously, 
 \begin{equation}
C_0:= \dmean{j_{0,\bv_1},j_{0,\bv_1}} =\sum_{x\in \Z^d} 
\left<j_{0,\bv_1} j_{x,x+\bv_1}\right>_0
\le \tfix \left< (V'(q_{e_1} - q_0))^2\right>_0
<\infty\, ,
 \end{equation}
and thus
$j_{0,\bv_1} \in \mathcal L^2$.  Then $u_\lambda\in L^2(\mu_{\tfix})$, and by stationarity
$\mean{u_\lambda}_0 = 0$.
We will show next that, in fact, $u_\lambda\in \mathcal H_1$.  
From \eqref{eq:res} we obtain
\begin{equation}
  \label{eq:ene}
  \lambda \mean{(\Phi_n u_\lambda)^2}_0
+ \mean{(\Phi_n u_\lambda) (-S)(\Phi_n  u_\lambda)}_0   =
  \mean{(\Phi_n u_\lambda) (\Phi_n j_{0,\bv_1})}_0  \, ,
\end{equation}
where we have used translation invariance of $L$ and antisymmetry of $A$.
Since $S j_{0,\bv_1} = -\gamma j_{0,\bv_1}$,  an application of Schwarz inequality yields
\begin{equation}
  \begin{split}
 \mean{(\Phi_n u_\lambda) (\Phi_n j_{0,\bv_1})}_0
    &=\ \gamma^{-1} \mean{(\Phi_n u_\lambda) (-S) (\Phi_n j_{0,\bv_1})}_0 \\
    &\le \gamma^{-1} \mean{(\Phi_n u_\lambda)(-S)(\Phi_n u_\lambda)}_0^{1/2}
 \mean{ (\Phi_n j_{0,\bv_1}) (-S) (\Phi_n j_{0,\bv_1})}_0^{1/2} \\
    &=  \gamma^{-1/2} \mean{ (\Phi_n j_{0,\bv_1})^2}_0^{1/2}
\mean{(\Phi_n u_\lambda)(-S)(\Phi_n u_\lambda)}_0^{1/2} \, .
  \end{split}
\end{equation}
Consequently, we have
\begin{equation}
  \label{eq:14}
\mean{(\Phi_n u_\lambda)(-S)(\Phi_n u_\lambda)}_0 \le \gamma^{-1} \mean{ (\Phi_n j_{0,\bv_1})^2}_0
\overset{n\to\infty}{\longrightarrow} \gamma^{-1} C_0\, ,
\end{equation}
which implies that
\begin{equation}
  \label{eq:ebb2}
     \lambda \dmean{u_\lambda, u_\lambda}  \le \gamma^{-1} C_0 
\end{equation}
and
\begin{equation}
  \label{eq:ebb}
      \| u_\lambda \|_1^2  \le \gamma^{-1} C_0  \, .
\end{equation}
Therefore, $u_\lambda\in \mathcal H_1$ and
by (\ref{eq:ebb}), we can extract a subsequence, which we still denote
with $u_\lambda$, weakly convergent in $\mathcal H_1$ to $u_0$.

Let $u_\lambda (p,q)= u_\lambda^s (p,q) + u_\lambda^a (p,q)$ where
 $u_\lambda^s$ and $u_\lambda^a$ are respectively symmetric and antisymmetric
in the $p$'s. Since $j_{0,\bv_1}$ is antisymmetric in the $p$'s, we have that 
$\dmean{u_\lambda , j_{0,\bv_1}}  = \dmean{ u_\lambda^a, j_{0,\bv_1}} $. 
Furthermore, $S$ preserves the parity in $p$, while it is inverted by $A$.
 So we can decompose the resolvent equation as
\begin{equation}
  \label{eq:sys}
  \begin{split}
    &\lambda u_\lambda^s - S u_\lambda^s - A u_\lambda^a = 0\, ,\\
    &\nu u_\mu^a - S u_\nu^a - A u_\nu^s = j_{0,\bv_1} \, . 
  \end{split}
\end{equation}
Taking a scalar product of the first equation with $u^s_\nu$,  of the second with
$u^a_\lambda$, and using the antisymmetry of $A$, we find
\begin{equation}\label{eq:sys2}
  \begin{split}
  \dmean{u^a_\lambda, j_{0,\bv_1}}  =  \nu \dmean{  u_\nu^a, u_\lambda^a }  +
  \dmean{ u_\lambda^a, (-S) u_\nu^a }  - \dmean{ u_\lambda^a, A u_\nu^s }  \\
  =  \nu \dmean{  u_\nu^a, u_\lambda^a }  + \lambda \dmean{ u_\lambda^s, u_\nu^s
  }  + \dmean{  u_\lambda, (-S) u_\nu } \, .
  \end{split}
\end{equation}

Since
\begin{equation}
  \int u_\lambda^a(p,q) \tilde{\mu}_{\tfix}(dp) = 0\, ,
\end{equation}
where $\tilde{\mu}_{\tfix}(dp)$ is the centered Gaussian product measure of
variance $\tfix$, and $S$ has a spectral gap $\gamma$ in $L^2(\tilde{\mu}_{\tfix}(dp))$, 
we have that
\begin{equation}
  \dmean{ u^a_\lambda, u^a_\lambda }  \ \le \ \frac{1}{\gamma}  
\dmean{ u_\lambda, (-S) u_\lambda } 
   \ \le\  C_0\gamma^{-2}\, .
\end{equation}
In particular, $u_0^a \in \mathcal L^2$.
Thus by taking first the limit as $\lambda\to 0$  we have 
$\lambda \dmean{ u_\lambda^s, u_\nu^s} \to 0$, then
 as $\nu\to 0$ we have $\nu \dmean{  u_\nu^a, u_0^a } \to 0$,
and finally we obtain from (\ref{eq:sys2})
\begin{equation}
   \dmean{u_0, j_{0,\bv_1}}  \ =\ \dmean{  u_0, (-S) u_0 } =   \| u_0\|_1^2\, .
\end{equation}
On the other hand, we have
\begin{equation}\label{eq:u01lim}
  \begin{split}
     \dmean{u_0, j_{0,\bv_1}}  \ =\  \lim_{\lambda\to 0} \dmean{u_\lambda,
     j_{0,\bv_1}} 
     = \lim_{\lambda \to 0} \left[ \lambda \dmean{ u_\lambda, u_\lambda } 
       + \dmean{  u_\lambda, (-S) u_\lambda } \right ] \\
     \ge  \lim_{\lambda \to 0} \lambda \dmean{ u_\lambda, u_\lambda }  +
     \| u_0\|_1^2 \, .
  \end{split}
\end{equation}
This implies
\begin{equation}
    \lim_{\lambda \to 0}\lambda \dmean{ u_\lambda, u_\lambda }  \ =\ 0 \, ,
\end{equation}
and
\begin{equation}
  \label{eq:15}
    \| u_\lambda\|_1 \to \| u_0 \|_1 \, .
\end{equation}
Therefore, $u_\lambda \to u_0$ strongly in $\mathcal H_1$. 

Uniqueness of the limit follows by the following standard argument. 
Suppose that $\lambda_n$ is the chosen subsequence such that $u_{\lambda_n}$ converges to $u_0$, and suppose $\nu_m$ is another sequence such that   $u_{\nu_m}$ converges to $\tilde u_0$. Then, similarly as we have done in equation \eqref{eq:sys2}
\begin{equation}
  \label{eq:16}
  \begin{split}
  \dmean{u^a_{\lambda_n}, j_{0,\bv_1}} 
  =  \nu_m \dmean{  u_{\nu_m}^a, u_{\lambda_n}^a }  + \lambda_n \dmean{ 
    u_{\lambda_m}^s, u_{\nu_n}^s
  }  + \dmean{  u_{\lambda_n}, (-S) u_{\nu_m} } 
  \end{split}
\end{equation}
which implies
\begin{equation}
   \dmean{u^a_{0}, j_{0,\bv_1}} = \dmean{  u_{0}, (-S) \tilde u_{0} } \, .
\end{equation}
Using $u^a_{\nu_m}$ instead of  $u^a_{\lambda_n}$, we find similarly
 \begin{equation}
   \dmean{\tilde u^a_{0}, j_{0,\bv_1}} = \dmean{  u_{0}, (-S) \tilde u_{0} } \, .
\end{equation}
Combining these with (\ref{eq:u01lim}) shows that $\norm{u_0-\tilde u_0}^2=0$, i.e.,
$u_0 = \tilde u_0$.

Thus the conductivity $\kappa(\tfix)$ defined by (\ref{eq:gk1}) is independent of the subsequence chosen for $\lambda$.
Moreover, we have 
\begin{equation}
 \kappa(\tfix) =
 \tfix^{-2} \dmean{u_0, j_{0,\bv_1}}  \ =\ \tfix^{-2}  \| u_0\|_1^2 \le \frac{C_0}{\tfix^2 \gamma} 
\le \frac{\left<V'(q_{e_1} - q_0)^2\right>_0}{\tfix \gamma}  \, .
\end{equation}
This completes the proof of Theorem \ref{infinite}.

\medskip

\section{Concluding remarks}
\label{sec:conclusions}

While all the results obtained in this paper are as expected, the difficulty
of actually proving things about the NESS of systems with nonlinear
dynamics is immense.  This is well illustrated by the impossibility (for
us) of obtaining a bound on the self-consistent temperature $T$ of the
second oscillator in a system consisting of three oscillators with
$T_1=\tlb$, $T_3=\trb$, and the Hamiltonian is as in (\ref{eq:1}) with
$\gamma_x = \gamma>0$.  We certainly expect that $T$ will satisfy
$\trb<T<\tlb$, but do not know how to prove this. All we know is that there
exists a $T=\mean{p_2^2}$, and that 
$\bar{j}= \tlb-\mean{p_1^2}=\mean{p_3^2}-\trb>0$.  We also know for
general $N$ that when
$\tlb, \trb\to \tfix$, then there is a self-consistent choice $T\to \tfix$,
and that this in this limit $(2N+1) \bar{j}/(\tlb-\trb)$ is bounded and
given by the Green-Kubo formula (\ref{eq:1stgGK}). 
Beyond this however we are stymied except when $V$ and $W$ are harmonic.
In that case $T$ is given by (\ref{eq:T-taylor}) without a correction term
for any 
$\tlb$, $\trb$, and due to explicit expressions $\gsc (x)$ can be analyzed in
great detail, proving $\trb<T<\tlb$.

\subsection*{Acknowledgment}

We thank Jonathan Mattingly and S.R.S. Varadhan for the help in the
proof of the existence of the self-consistent profile.
We also thank Herbert Spohn for useful discussions.  
The work of F.~Bonetto was
supported in part by NSF Grant DMS-060-4518, the work of
J.~L.~Lebowitz was supported in part by NSF Grant DMR-044-2066 and by
AFOSR Grant AF-FA 9550-04-4-22910, the work of J.~Lukkarinen by
Deutsche For\-schungs\-gemein\-schaft (DFG) project SP~181/19-2 and by
the Academy of Finland, the work of S.~Olla by 
ANR LHMSHE no.\ BLAN07-2 184264 (France).

\end{document}